\documentstyle[preprint,tighten,aps]{revtex}
\begin{document}
\preprint{\vbox{ \null\hfill INFNCA-TH9514 \\
\null\hfill LPC 95 24 \\
\null\hfill hep-ph/9511432}}
\draft
\title{ Glueball plus pion production in hadron collisions}

\author{ Francesco Murgia}

\address{Istituto Nazionale di Fisica Nucleare,
Sezione di Cagliari \\
via Ada Negri 18, I--09127 Cagliari, Italy }
%({\tt e-mail: murgia@cagliari.infn.it; tel: +39 70 670834;\\
%fax: +39 70 657823})}

\author{ Paul Kessler and Joseph Parisi}

\address{Laboratoire de Physique Corpusculaire, Coll\`ege de France \\
11, Place Marcelin Berthelot F-75231 Paris Cedex 05, France }
%({\tt e-mail: parisi@frcpn11.in2p3.fr; tel: +33 1 44271456;\\
%fax: +33 1 43546989})}

\date{November 1995}

\maketitle

\begin{abstract}
Using a non--relativistic gluon bound--state model for glueballs (G),
we compute the subprocess $q\,\bar q\, \to\, G\,\pi$, and we therefrom
derive the yield of the overall reaction $p\,\bar p\, \to\, G\,\pi X$,
assuming the glueball and the pion to be emitted with their
transverse momenta large, opposite and approximately equal.
Numerical results are presented in the form of $p_T$ spectra
for various glueball candidates and their possible quantum states,
assuming those particles to be produced, in the type of reactions
here considered, at high--energy $p\,\bar p$ colliders such as the
CERN Sp\=pS.
\end{abstract}


\narrowtext

\renewcommand{\thefootnote}{\fnsymbol{footnote}}

\section{Introduction}
\label{intro}

A non--relativistic gluon bound--state model for computing the
production and decay of glueballs (G) made of two gluons was proposed
a few years ago by Kada {\it et al.} \cite{kada}, who used it
in order to calculate the processes $J/\psi\, \to\, G\,\gamma$ and
$G\, \to\, \gamma\,\gamma$. That model was later generalized for
more complex reactions by Houra--Yaou {\it et al.} \cite{hour},
and applied by them to the production of a glueball plus
a quark or gluon jet at high transverse momentum in hadron
collisions. Another application, recently computed by
Ichola and Parisi \cite{icho}, concerned glueball plus pion
production in two--photon processes. In this paper we consider
the production of the same final state as in Ref.~\cite{icho},
but this time in hadron collisions.

Indeed, while the existence of glueballs is considered a crucial
test of quantum chromodynamics \cite{frit}, and after a few
glueball candidates have emerged in the early eighties from
various experiments \cite{pala}, further experimental evidence
appears still necessary in order to firmly establish their
nature and properties. Besides other reactions that should
involve a ``gluon--rich environment'' (such as radiative $J/\psi$
decays, as well as diffractive hadron--hadron scattering assumed
to involve double Pomeron exchange), hard collisions occurring
in high--energy reactions may provide another means of
creating that kind of environment and thus producing glueballs.

We are aware that the status of the three particles that were
considered as glueball candidates in Refs.~\cite{kada,hour,icho},
namely the $\eta(1440)$, the $f_2(1720)$, and the $X(2220)$,
has become more uncertain in the last few years
\cite{heus,eige,pdg4}. However, as has been discussed
at large in Ref.~\cite{icho}, recent experimental data
regarding those particles are rather contradictory;
actually there has also been recently some positive evidence
regarding the $f_2(1720)$ \cite{pal2} and the $X(2220)$
\cite{chao}. Anyway, for none of the three candidates
it has been decisively proved that it should not be a glueball.
Therefore, in this paper, we still stick to the assumptions of
Refs.~\cite{kada,hour,icho}.

Hereafter, in section II, we recall the formalism used and
present the details of our calculation. Section III contains
a discussion of the numerical results obtained and a brief
conclusion. Two appendices provide respectively the
expressions of all quantities (four--momenta, polarization
four--vectors, projectors of spinor pairs) needed for our
calculation and those of all independent helicity
amplitudes obtained for the subprocess
$q\,\bar q \,\to\, G\,(q'\bar q')_{PS}$ resp. $q\,\bar q \,\to\, G\,\pi$.

\section{Description of the formalism and details of calculation}
\label{descri}

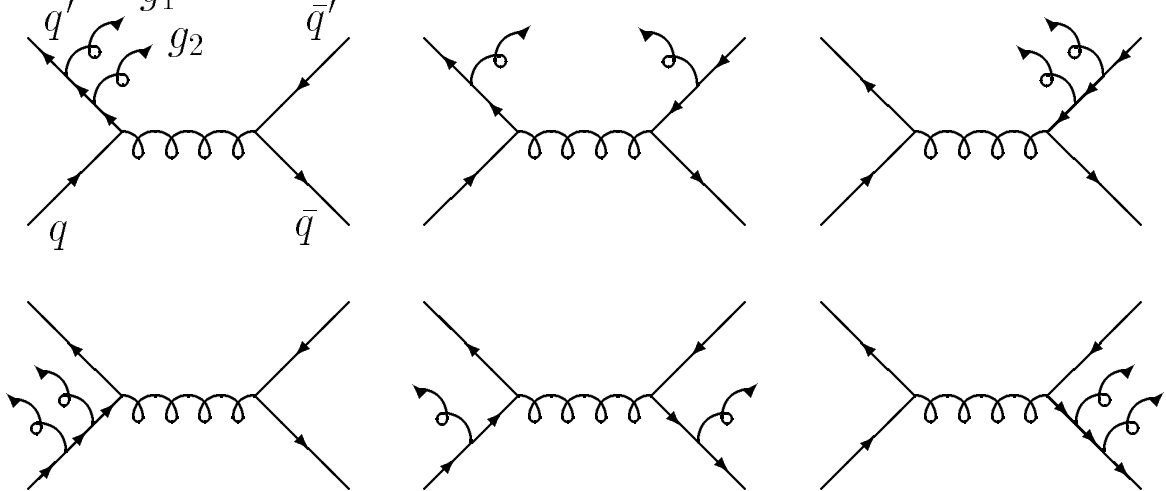
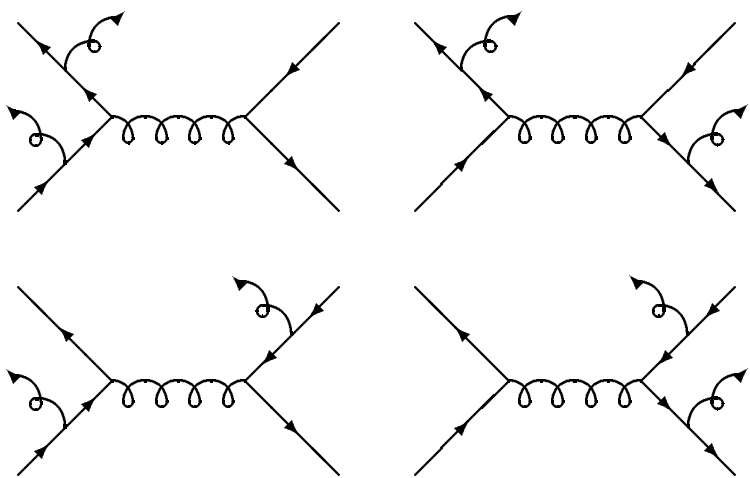
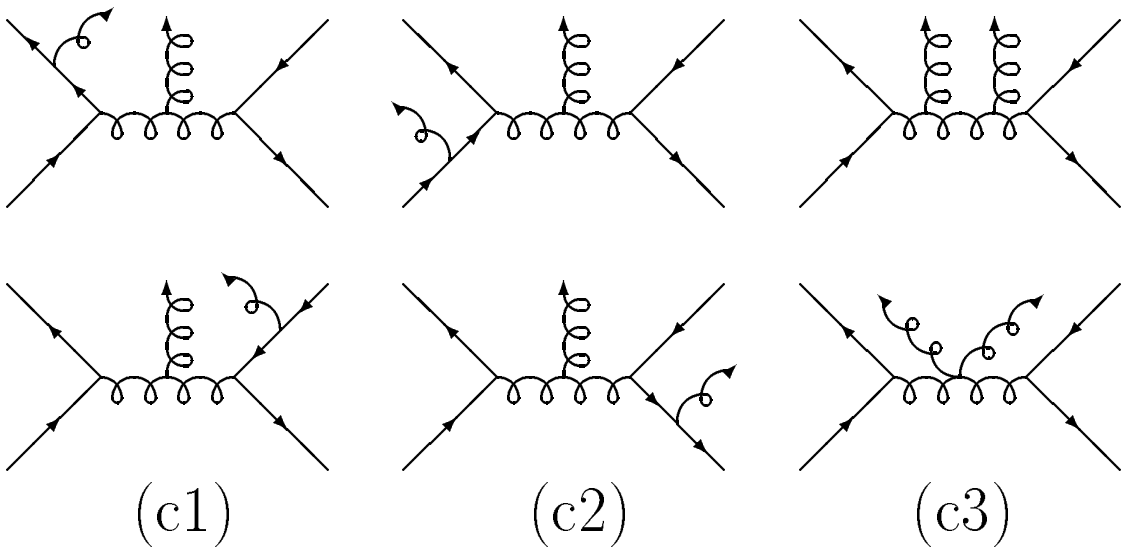
\begin{figure}
\caption[fig1]{ The Feynman graphs that, to lowest order in QCD,
contribute to the process $q\,\bar q \,\to\, g_1\,g_2\,q'\,\bar q'$.
The graphs are grouped according to their color factors
(a convolution with the color part of the glueball
and the pion wave function is understood):
a) $c_F = 8\delta_{ij}\protect/(9\sqrt{6})$;
b) $c_F = -\,\delta_{ij}/(9\sqrt{6})$;
c1) $c_F = i\,\delta_{ij}/\sqrt{6}$;
c2) $c_F = -i\,\delta_{ij}/\sqrt{6}$;
c3) $c_F = -\,2\delta_{ij}/\sqrt{6}$.
$i$, $j$ are the color indices of the incoming quark and
antiquark, respectively. For each graph, except for that
including the four--gluon vertex, the corresponding one
where $g_1$ and $g_2$ are exchanged must be also taken into
account. }
\label{graphs}
\end{figure}

Let us first remark that, for $G\,\pi$ production in hadronic
reactions, the hard process is necessarily induced by
quark--antiquark collisions. Indeed, the subprocess
$g\,g \,\to\, G\,\pi^0$ is excluded since,
due to isospin conservation,
the pion cannot be coupled to any parton system composed
exclusively of gluons. For the same reason, the
subprocess $q\,\bar q \,\to\,  G\,\pi^0$ cannot involve any
Feynman diagram where the quark and antiquark annihilate
into a gluon. Therefore the calculation of the hard
subprocess is the same for $G\,\pi^0$ and $G\,\pi^\pm$
production; at lowest order in perturbative QCD it involves
the diagrams shown in Fig.~\ref{graphs}, where the gluons $(g_1,g_2)$
and the final quarks $(q',\bar q')$ are, respectively,
the components of the glueball and the pion to be
produced.

Applying the non--relativistic gluon bound--state model
\cite{kada,hour,icho} for glueballs, together with
the well known Brodsky--Lepage model \cite{brod}
for pions, we write, in analogy with Eq.~(1) of
Ref.~\cite{icho}:

\begin{eqnarray}
{\cal M}^{\lambda\,\bar\lambda,\,\Lambda}_{q\,\bar q \,\to\,
G\,(q'\bar q')_{PS}}(E,\Theta,z) & = &
f_L \lim_{\beta\to 0} \frac{1}{\beta^L}
\int \frac{d(\cos\theta)\,d\varphi}{4\pi} \nonumber \\
& \times & \sum_{\lambda_1,\lambda_2} \zeta^{LSJ\Lambda}
_{\lambda_1\lambda_2}(\theta,\varphi)\,{\cal M}^{\lambda
\,\bar\lambda,\,\lambda_1\,\lambda_2}_{q\,\bar q\,\to\,
g_1\,g_2\,(q'\bar q')_{PS}}(E,\Theta,\theta,\varphi,z)
\label{bound}
\end{eqnarray}

\noindent where we have used the following definitions:
$E$ and $\Theta$ are, respectively, the total energy
and the pion emission angle in the $q\,\bar q$
center--of--mass frame, while $\theta$ ($\varphi$)
is the polar (azimuthal) emission angle of either gluon
in the $g_1\,g_2$ c.m. frame, {\it i.e.} the glueball
rest frame (see Fig.~\ref{frame}). $z$ is the Brodsky--Lepage
variable defining the fractional momentum of the quark $q'$
within the pion. We call $J$, $L$, $S$, $\Lambda$ respectively
the total spin of the glueball, its orbital angular momentum,
its intrinsic spin, and the component of its total spin
along the $z$--axis of Fig.~\ref{frame}. In addition, we
call $\lambda$ ($\bar\lambda)$ the helicities of $q$
($\bar q$), while $\lambda_1$ ($\lambda_2$) are the
helicities of $g_1$ ($g_2$) in the glueball rest frame.
The angular projection function
$\zeta^{LSJ\Lambda}_{\lambda_1\lambda_2}(\theta,\varphi)$
is defined as

\begin{equation}
\zeta^{LSJ\Lambda}_{\lambda_1\lambda_2}(\theta,\varphi) =
d^J_{\Lambda\bar\Lambda}(\theta) \,e^{-i\Lambda\varphi}
\langle L\ S\ 0\ \bar\Lambda\mid L\ S\ J\ \bar\Lambda\rangle
\langle 1\ 1\ \lambda_1,-\lambda_2\mid 1\ 1\ S\ \bar\Lambda\rangle
\label{zeta}
\end{equation}

\noindent where $\bar\Lambda = \lambda_1-\lambda_2$. $\beta$
is the velocity of either gluon in the glueball rest frame,
while $f_L$ is given by

\begin{equation}
f_L = \sqrt{\frac{2L+1}{2\pi M}}\left(-\frac{2i}{M}\right)^L
\frac{(2L+1)!!}{L!}\left[\left(\frac{d\,\ }{dr}\right)^L
R_L(r)\right]_{r=0}
\label{fl}
\end{equation}

\noindent where $M$ is the glueball mass and $R_L(r)$
its radial wave function in configuration space.
Finally we notice that the system $q'\bar q'$ is here
assumed to be in a pseudoscalar (PS) state.

In the following stage we apply the Brodsky--Lepage
convolution formula \cite{brod}:

\begin{equation}
{\cal M}^{\lambda\,\bar\lambda,\,\Lambda}_{q\,\bar q\,\to\,
G\,\pi}(E,\Theta) = \int dz\,\Phi_{\pi}(z)\,{\cal M}
^{\lambda\,\bar\lambda,\,\Lambda}_
{q\,\bar q\,\to\, G\,(q'\bar q')_{PS}}(E,\Theta,z)
\label{conv}
\end{equation}

\noindent where $\Phi_{\pi}(z)$ is the pion distribution
amplitude.

As in Refs.~\cite{hour,icho} we assume the glueball to be
extreme--relativistic in the $q\bar q$ c.m. frame, {\it i.e.}
$M/E \to 0$. In that approximation the gluons are also
treated as massless in the hard subprocess. {\it A fortiori}
the mass of the pion, as well as of its constituent quarks,
is also neglected. In other words: both outgoing particles,
and all partons involved, are on the light cone.

It is to be noticed that, with massless quarks, helicity
conservation \cite{brod} imposes: $\bar\lambda = - \lambda$.
On the other hand, due to parity and angular--momentum
conservation, one has the relation ${\cal M}^{-\lambda,
-\bar\lambda,-\Lambda}_{q\,\bar q\,\to\, G\,\pi} =
(-1)^{J+L+\Lambda+1}{\cal M}^{\lambda\,\bar\lambda,\,\Lambda}
_{q\,\bar q\,\to\, G\,\pi}$, which reduces the number of independent
amplitudes by an additional factor of two. It thus becomes
sufficient to limit oneself to computing those amplitudes
where $\lambda = 1/2$, $\bar\lambda = -1/2$.

In Appendix A we show the expressions of four--momenta,
polarization four--vectors and spinors needed for our calculation.
Appendix B contains the expressions of the independent
helicity amplitudes obtained, corresponding to the various
glueball quantum states considered, both after applying
Eq.~(\ref{bound}) and after we use Eq.~(\ref{conv}) involving
a convolution with the pion distribution amplitude
$\Phi_{\pi}(z)$. For the latter we choose two different
expressions, namely that proposed by Chernyak and
Zhitnitsky \cite{cher}:

\begin{equation}
\Phi^{CZ}_{\pi}(z) = 5\sqrt{3}f_\pi\,z(1-z)(2z-1)^2
\label{czda}
\end{equation}

\noindent and the so--called asymptotic one \cite{brod}

\begin{equation}
\Phi^{as}_{\pi}(z) = \sqrt{3}f_\pi\,z(1-z)
\label{asda}
\end{equation}

\noindent where $f_\pi$ is the pion decay constant
($f_\pi \cong 93$ MeV).

{}From the amplitudes thus obtained one derives the
transverse--momentum spectrum for the subprocess considered,
taking account of kinematic factors (where again one makes
$M/E \to 0$):

\begin{eqnarray}
\frac{d\sigma^{q\bar q\to G\pi}}{dp_T}(E,p_T) & = &
\frac{p_T}{288\pi E^3 \sqrt{E^2-4p_T^2}} \nonumber \\
& \times & \sum_{i,j}\sum_{\lambda,\bar\lambda,\Lambda}
\left[\mid{\cal M}^{\lambda\,\bar\lambda,\,\Lambda}_{q\,\bar q
\,\to\, G\,\pi}(E,\Theta)\mid^2 +
\mid{\cal M}^{\lambda\,\bar\lambda,\,
\Lambda}_{q\,\bar q\,\to\, G\,\pi}
(E,\pi-\Theta)\mid^2\right]
\label{dspar}
\end{eqnarray}

\noindent where, in the expressions of the amplitudes,
$\cos\Theta$ is to be replaced by $(1-4p_T^2/E^2)^{1/2}$
and $\sin\Theta$ by $2p_T/E$; $i,j$ are the color indices of
$q,\bar q$ respectively.

The transverse--momentum spectrum for the overall reaction
$p\bar p \to G\pi X$ is then given by convoluting the
spectrum defined by Eq.~(\ref{dspar}) with the distribution
functions of the quarks and the antiquarks inside the proton and
the antiproton, as follows:

\begin{equation}
\frac{d\sigma^{p\bar p\to G\pi X}}{dp_T}(s,p_T) =
\int_{x_{min}}^1 \!\!\! dx \int_{x'_{min}}^1 \!\!\! dx'
F(x,x',\mbox{``}Q^2\mbox{''})
\frac{d\sigma^{q\bar q\to G\pi}}{dp_T}(E,p_T)
\label{dsadr}
\end{equation}

\noindent where $s$ is the overall c.m. energy squared;
noticing that $E^2=xx's$, one gets: $x'_{min}=E^2_{min}/(xs)$,
$x_{min} = E^2_{min}/s$, with $E_{min}=2p_T$.
As for the function $F(x,x',\mbox{``}Q^2\mbox{''})$, it
is defined in the following way:

\begin{description}
\item[(i)] For $G\,\pi^\pm$ production

\begin{equation}
F(x,x',\mbox{``}Q^2\mbox{''}) =
f_{u/p}(x,\mbox{``}Q^2\mbox{''})
f_{d/p}(x',\mbox{``}Q^2\mbox{''}) +
f_{\bar d/p}(x,\mbox{``}Q^2\mbox{''})
f_{\bar u/p}(x',\mbox{``}Q^2\mbox{''})
\label{fchar}
\end{equation}

\item[(ii)] For $G\,\pi^0$ production

\begin{eqnarray}
F(x,x',\mbox{``}Q^2\mbox{''}) & = & \frac{1}{2}\Bigl[
f_{u/p}(x,\mbox{``}Q^2\mbox{''})
f_{u/p}(x',\mbox{``}Q^2\mbox{''}) +
f_{d/p}(x,\mbox{``}Q^2\mbox{''})
f_{d/p}(x',\mbox{``}Q^2\mbox{''}) \nonumber \\ & + &
f_{\bar u/p}(x,\mbox{``}Q^2\mbox{''})
f_{\bar u/p}(x',\mbox{``}Q^2\mbox{''}) +
f_{\bar d/p}(x,\mbox{``}Q^2\mbox{''})
f_{\bar d/p}(x',\mbox{``}Q^2\mbox{''}) \Bigr]
\label{fzero}
\end{eqnarray}

\end{description}

Here we have made use of the fact that the quark (antiquark)
content of the antiproton is equal to the antiquark (quark)
content of the proton.

For the distribution functions $f_{q/p}$, $f_{\bar q/p}$
we use the parametrization CTEQ3 (leading order QCD)
\cite{cteq}, while for the scale parameter we take
$\mbox{``}Q^2\mbox{''} = M^2$.

In order to eliminate the normalization constant $f_L^2$
(see Eq.~(\ref{bound})), we use the same procedure as
in Refs.~\cite{kada,hour,icho}, {\it i.e.} we write:

\begin{equation}
\frac{d\sigma^{p\bar p\to G\pi X}}{dp_T} B(G\to x\,y\dots) =
\frac{d\sigma^{p\bar p\to G\pi X}}{dp_T}
\frac{\Gamma(J/\psi\to G\,\gamma)B(G\to x\,y\dots)}
{\Gamma(J/\psi\to G\,\gamma)}
\label{ratio}
\end{equation}

\noindent where $B(G\to x\,y\dots)$ is the branching ratio for
glueball decay in a given channel (actually we shall
consider only the main decay channel for each glueball
candidate).
Then the numerator in the second factor on the right--hand
side of Eq.~(\ref{ratio}) is given by experimental
measurements, while for the corresponding denominator
we use the expressions computed before \cite{kada}\footnote{
The decay widths given in Ref.~\protect\cite{kada}
have been systematically multiplied by a factor
of 4, since the helicity amplitudes had been
underestimated there by a factor of 2. In
addition two misprints that appeared there,
regarding the widths of $J/\psi$ radiative
decay into glueballs with quantum states
$J=0$, $L=S=1$ and $J=4$, $L=S=2$, have
been corrected. Furthermore we have slightly
modified the values of the numerator of the
second factor on the r.h. side of Eq.~(\ref{ratio}),
in accordance with the most recent experimental
data (see Refs.~\protect\cite{pdg4,chao});
in addition,
for the $\eta(1440)$, we have here considered
its decay in the $\rho^0\rho^0$ (instead of
$K\bar K\pi$) channel, and for the $X(2220)$
its decay in the $\pi^+\pi^-$ (instead of
$K\bar K$) channel.}.

%(see however \cite{not1}).

Then, in principle, there is no free parameter left; yet
there is a certain freedom of choice regarding the
expressions of the $\alpha_s$ factors present in the
calculation. Notice that on the r.h. side of
Eq.~(\ref{ratio}) we get a factor $\alpha_s^4$
(coming from the helicity amplitudes, see
Eqs.~(\ref{dspar}),(\ref{dsadr}) and appendix B),
divided by a factor $\alpha_s^2$ contained in the
$J/\psi$ partial decay width. Assuming that $\alpha_s$
takes approximately the same value in both processes
considered, {\it i.e.} $p\,\bar p \,\to\, G\,\pi\, X$ and
$J/\psi\,\to\, G\,\gamma$, we are left with a factor $\alpha_s^2$
in the final expression of the transverse--momentum spectra.
We take: $\alpha_s(\mbox{``}Q^2\mbox{''}) =
\alpha_s(M^2) = 12\pi/[25\ln(M^2/\Lambda^2)]$, with
$\Lambda = 0.2$ GeV.

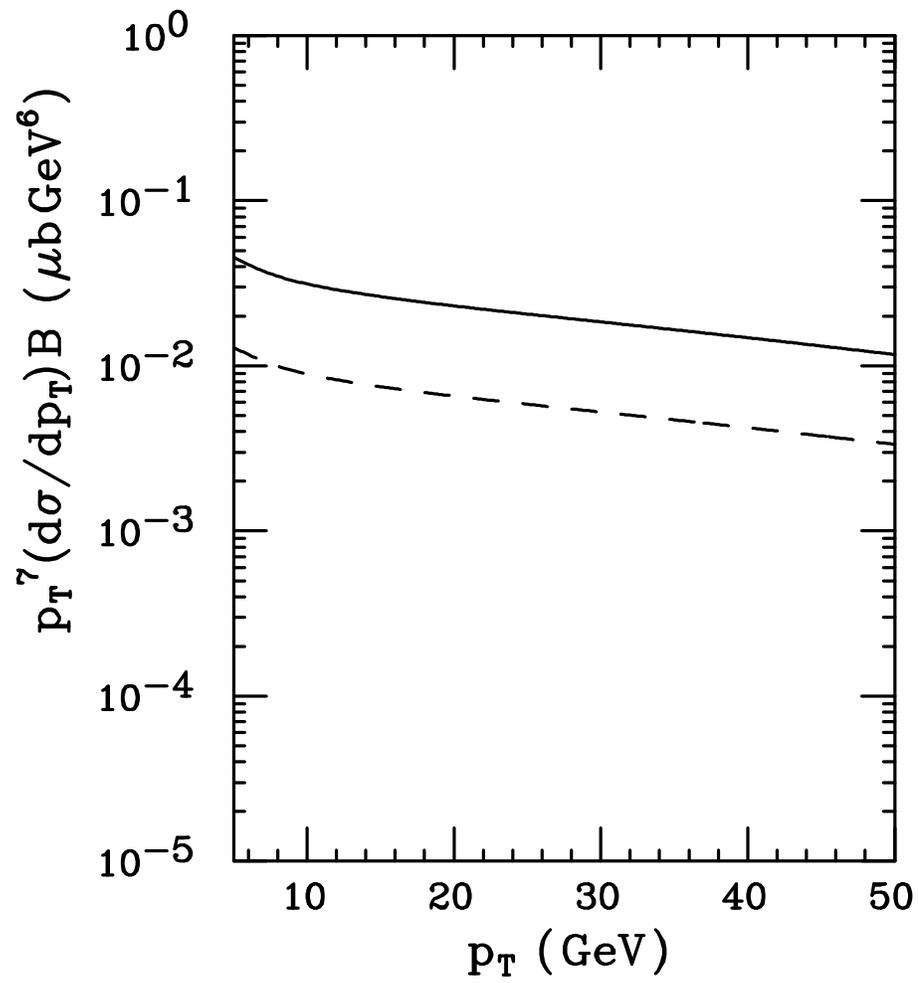
\begin{figure}
\caption[fig3]{ The transverse--momentum spectrum, multiplied by
$p_T^{\ \,7}$ and $B$, predicted for the reaction
$p\,\bar p \,\to\, G\,\pi^0\,X$ at $s = 4\times 10^5$ GeV$^2$.
Here $G = \eta(1440)$ and $B = Br(\eta(1440)\to \rho^0\rho^0)$.
Both the asymptotic (dashed curve) and the
Chernyak--Zhitnitsky (full curve) pion distribution
amplitudes have been considered.
The parametrization CTEQ3 (at leading order in QCD)
\protect\cite{cteq}
for the parton distribution functions has been used. }
\label{eta0}
\end{figure}

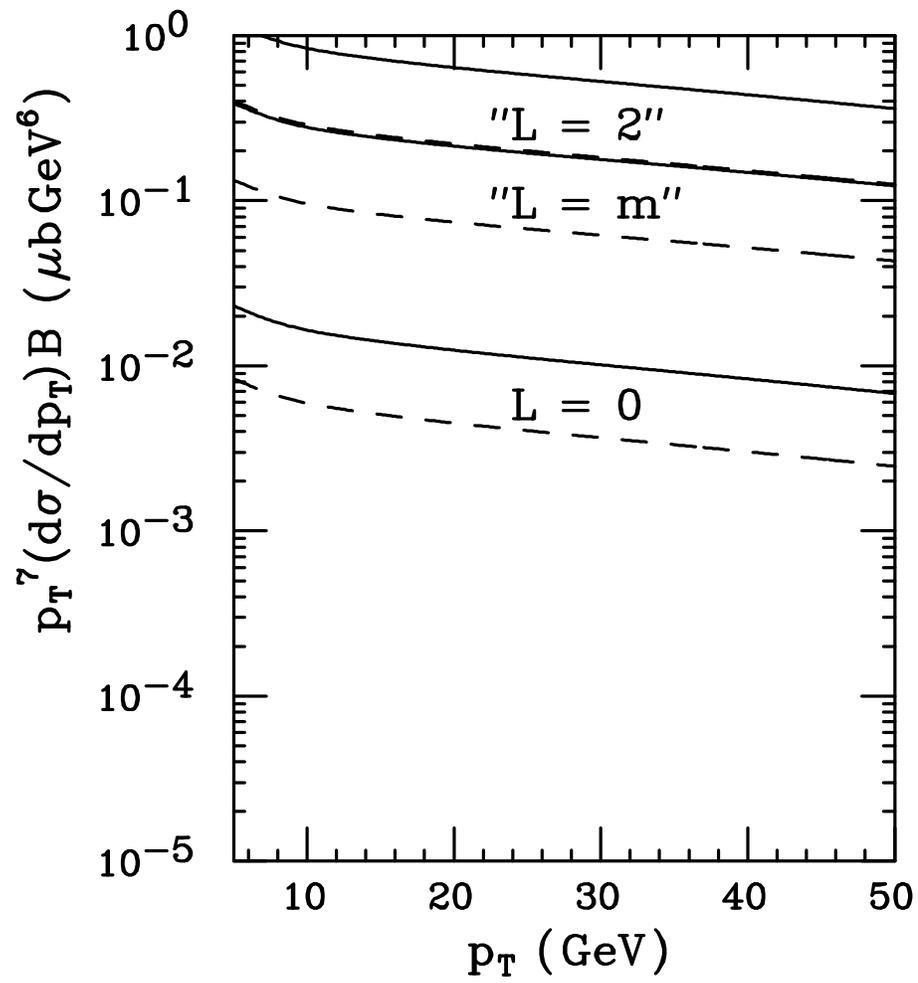
\begin{figure}
\caption[fig4]{ Same as Fig.~\protect\ref{eta0}, with $G = f_2(1720)$,
$B = Br(f_2(1720) \to K\bar K)$.
%The quantum states
%considered are of the form $|J=2\rangle = A_{02}|L=0,S=2;J=2\rangle
%+ A_{20}|L=2,S=0;J=2\rangle+ A_{22}|L=2,S=2;J=2\rangle$.
%In detail: a) $A_{02}=1$, $A_{20}=A_{22}=0$;
%b) $A_{20}=
The same quantum
states as in Refs.~\protect\cite{kada,hour,icho} are considered
for the $f_2(1720)$.}
\label{f2}
\end{figure}

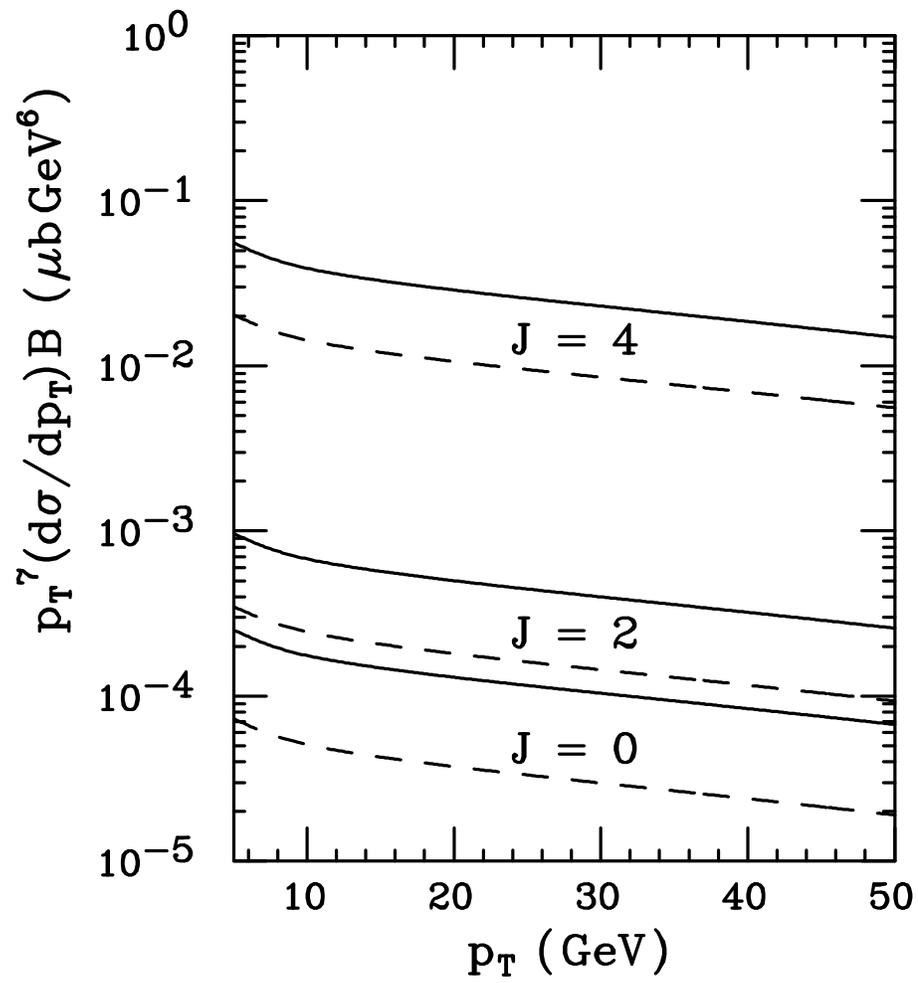
\begin{figure}
\caption[fig5]{ Same as Fig.~\protect\ref{eta0}, with $G = X(2220)$,
$B = Br(X(2220) \to \pi^+\pi^-)$. The same quantum
states as in Refs.~\protect\cite{kada,hour,icho} are considered
for the $X(2220)$. }
\label{csi}
\end{figure}

The $p_T$ spectra thus obtained for the reaction
$p\,\bar p\,\to\, G\,\pi^0\, X$ are shown, for the three
glueball candidates and their respective quantum states
here considered (the same as in Ref.~\cite{kada,hour,icho},
apart from a slight numerical modification\addtocounter{footnote}{3}
\footnote{The state called ``$L=m$'' in
Refs.~\protect\cite{kada,hour,icho}
is a mixture of states $L=0$,
$S=2$ and $L=2$, $S=0$ with their respective
weight coefficients connected by the relation
$A_{20}=0.286 M^2 [R_0(0)/R''_2(0)]A_{02}$,
while the state called ``$L=2$'' is a mixture
of states $L=2$, $S=0$ and $L=2$, $S=2$ with
their relative weight coefficients related by
$A_{20}=-6.816A_{22}$. Both mixtures have been
adjusted in such a way that they fit the
experimental ratios of helicity amplitudes
measured in the process $J/\psi \to \gamma
f_2(1720)$.}),
with either pion distribution amplitude assumed
(Eqs.~(\ref{czda}),(\ref{asda})), at an $s$ value of
$4\cdot 10^5$ GeV$^2$ (chosen so as to be approximately
that of the CERN Sp\=pS collider) in
Figs.~\ref{eta0},\ref{f2},\ref{csi}.

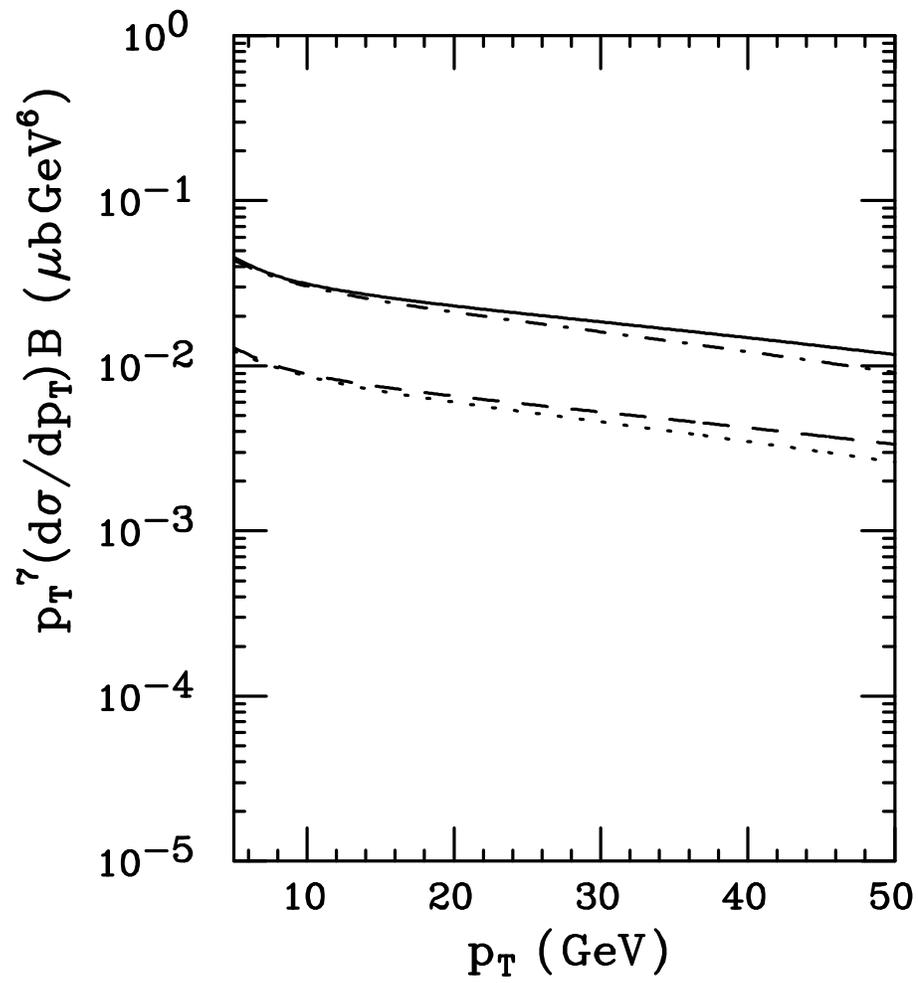
\begin{figure}
\caption[fig6]{ Comparison between the results of Fig.~\protect\ref{eta0}
and the corresponding ones for the reaction
$p\,\bar p \,\to\, G\,\pi^\pm\,X$.
Same notation as in Fig.~\protect\ref{eta0}. In addition,
the dot--dashed and dotted curves refer to $\pi^{\pm}$ production,
using respectively the CZ and asymptotic distribution amplitude. }
\label{etac}
\end{figure}

As is shown by Fig.~\ref{etac}, taking the case of the $\eta(1440)$
as an example, there is very little difference between
the respective $p_T$ spectra of neutral and charged
pions produced together with glueballs; the latter
are slightly lower.

\section{Discussion and conclusion}
\label{concl}

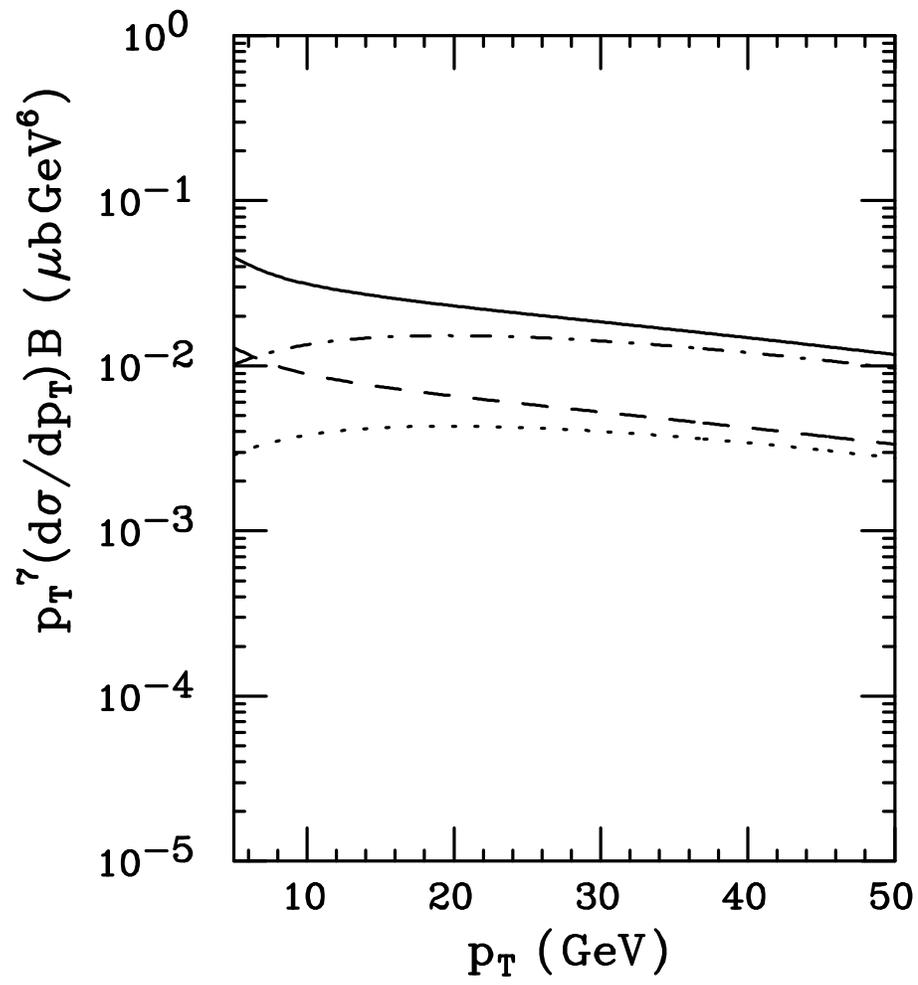
\begin{figure}
\caption[fig7]{ Comparison between the results of Fig.~\protect\ref{eta0} and
the analogous ones, obtained using a different set (MRSA)
of parton distribution functions (see Ref.~\protect\cite{mrsa}).
Same notation as in Fig.~\protect\ref{eta0}. In addition,
the dot--dashed and dotted curves correspond to MRSA,
using respectively the CZ and asymptotic distribution amplitude. }
\label{etamr}
\end{figure}

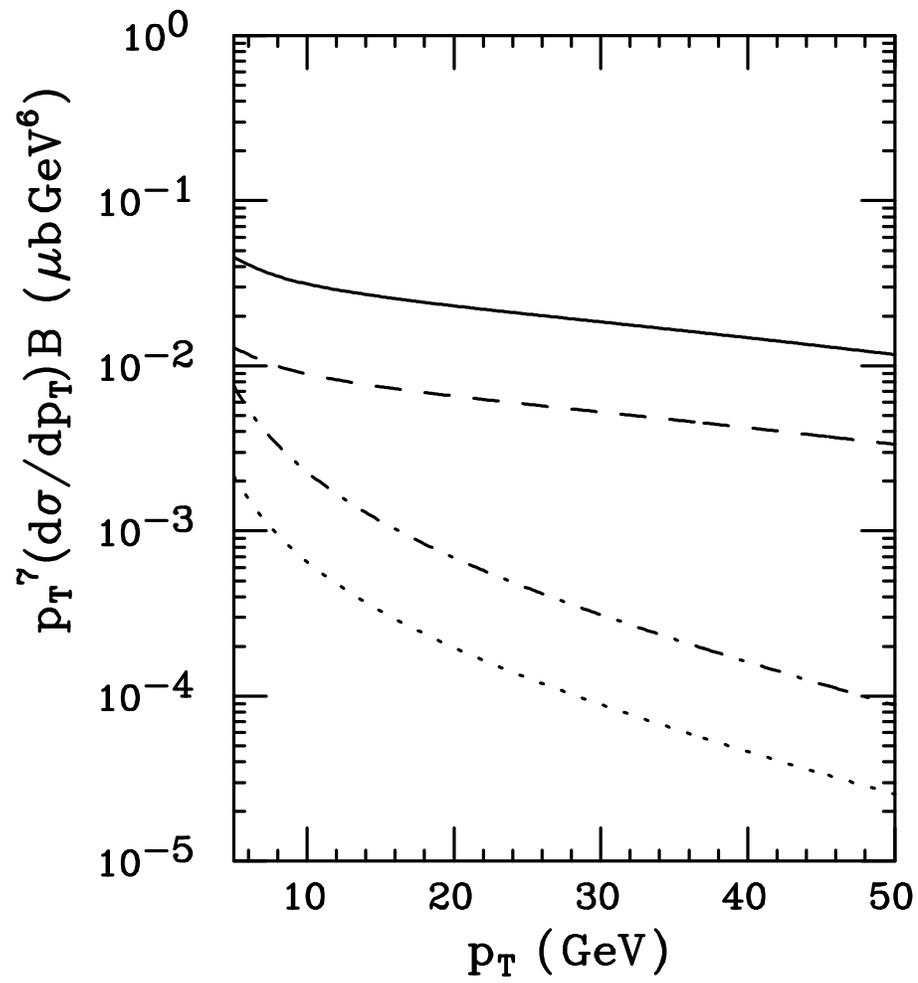
\begin{figure}
\caption[fig8]{ Comparison between the results of Fig.~\protect\ref{eta0}
and the analogous ones, obtained using a different prescription
for the value of ``$Q^2$'' in the expression of both the
function $F(x,x',\mbox{``}Q^2\mbox{''})$ (see Eq.~(\ref{dsadr}))
and $\alpha_s$.
Same notation as in Fig.~\protect\ref{eta0}. In addition,
the dot--dashed and dotted curves correspond to
``$Q^2$''$= p_T^2$ instead of ``$Q^2$''$= M^2$,
using respectively the CZ and asymptotic distribution amplitude. }
\label{etaq}
\end{figure}

Our results call for the following remarks:

\begin{description}
\item[(i)] The curves shown in
Figs.~\ref{eta0},\ref{f2},\ref{csi} are approximately flat,
{\it i.e.} the $p_T$ spectra obtained roughly scale like
$p_T^{-7}$, as could be predicted from dimensional
counting rules. Violations of that scaling rule are only
logarithmic.
\item[(ii)] Correspondingly, it is only through logarithmic
factors (of the type $\ln[s/(4p_T^2)]$ or $\ln^2[s/(4p_T^2)]$)
that those $p_T$ spectra depend on the machine energy. It results,
as we have checked, that there is a slight increase,
accompanied by a flattening of the curves, when one passes
from Sp\=pS to Tevatron energy, {\it i.e.} from
$s = 4\cdot 10^5$ GeV$^2$ to $s = 4\cdot 10^6$ GeV$^2$.
\item[(iii)] As usual, the yields predicted with the
Chernyak--Zhitnitsky distribution amplitude are
somewhat higher (by a factor of 3--4) than those computed
with the asymptotic one.
\item[(iv)] If one integrates the spectra over $p_T$
from $p_{T_{min}} = 5$ GeV on (assuming that there are
no additional drastic acceptance cuts), the integrated
cross sections obtained are of the order of $10^{-35}$
to $10^{-39}$ cm$^2$, depending on the glueball
candidate and quantum state considered, as well as
on the pion distribution amplitude chosen.
Some of those cross sections, {\it i.e.} those corresponding
to the $\eta(1440)$, to the states ``$L=2$'' and ``$L=m$''
of the $f_2(1720)$ and to the state $J=4$ of the
$X(2220)$, might be measurable under present experimental
conditions. This conclusion calls however for some
reservations, if the sources of uncertainty listed
hereafter in (v)--(vii) are taken into account.
\item[(v)] In our calculation we have retained only
lowest order terms in both the series expansion in
powers of $M/E$ and that in powers of $\alpha_s$.
Noticing that $M/E$
$\raisebox{+.1truecm}{$<$}$\hskip-0.32truecm
$\raisebox{-.1truecm}{$\sim$}$
$0.2$
(since $E_{min} = 2p_{T_{min}} = 10$ GeV) and
$\alpha_s(M^2) = 0.30\div 0.35$, it still seems reasonable
to expect that the inclusion of higher--order corrections
would not modify the orders of magnitude obtained.
\item[(vi)] The parametrization of the quark distribution
functions at very small values of $x, x'$ (here
$x_{min} = x'_{min} = 2.5\cdot 10^{-4}$) has not yet
been verified experimentally. As is shown in Fig.~\ref{etamr},
a different parametrization such as MSRA \cite{mrsa}
would lead to different shapes of the $p_T$ spectra
and consequently to significantly lower values
of the integrated cross sections.
\item[(vii)] One may also change the scale ``$Q^2$''
both in the expressions of the distribution functions
and in that of the $\alpha_s$ factors involved in the
calculation of $q\,\bar q \,\to\, G\,\pi$. For instance,
setting ``$Q^2$'' = $p_T^2$ instead of $M^2$ modifies
the $p_T$ spectra rather drastically and thus sharply
affects, as well, the integrated cross sections
(see Fig.~\ref{etaq}).
\item[(viii)] It is interesting to compare the yields here
obtained with those of other exclusive channels in
hadronic reactions, such as for instance pion pair
production in $p\,\bar p$ collisions. That process has been
computed, using the Brodsky--Lepage model \cite{brod},
by Djagouri et al. \cite{dja} who have however restricted
their calculation to a scattering angle of $90^\circ$ in
the c.m. frame of the hard subprocess. Comparing
$[d\sigma/d(\cos\Theta)]_{\Theta=90^\circ}$ for
$p\,\bar p \,\to\, G\,\pi^0\,X$ and
$p\,\bar p \,\to\, \pi^0\,\pi^0\,X$ under
the same experimental conditions ($s=4\cdot 10^5$ GeV$^2$,
$p_T > 5$ GeV) and with the same theoretical ingredients
(CTEQ3 parametrization, ``$Q^2$''$=M^2$), we get the following
result: The largest of the $G\,\pi^0$ production cross sections,
namely that for $G = f_2(1720)$ with ``$L=2$'', is about
one order of magnitude smaller than that obtained
for $\pi^0\,\pi^0$ production; more precisely,
the corresponding ratio is 0.42 with the asymptotic
pion distribution amplitude, and 0.063 with the CZ one.
Comparing, on the other hand, the reactions $p\,\bar p
\,\to\, G\,\pi^\pm\, X$ and $p\,\bar p \,\to\, \pi^+\,\pi^-\, X$,
the ratios of the corresponding yields are about half
of those obtained in the previous case.
Finally, comparing $p\,\bar p \,\to\, G\,\pi^0\, X$ or
$p\,\bar p \,\to\, G\,\pi^\pm\, X$ (still for the same choice of G)
with $p\,\bar p \,\to\, \pi^\pm\,\pi^0\, X$, the cross sections
computed are roughly of the same order (it is to be noticed
that in both reactions no gluon--gluon interaction
does contribute).
\item[(ix)] As compared with $p\,\bar p \,\to\, G\,\pi^0\, X$,
the reactions $p\,\bar p \,\to\, G\,\eta\, X$ and
$p\,\bar p \,\to\, G\,\eta'\, X$ would certainly be
more promising from a quantitative point of view,
since they would involve the contribution of gluon--gluon
interactions; that contribution may indeed be expected
to increase the $p_T$ spectra and the corresponding
integrated cross sections by several orders of magnitude
(see \cite{hour}).
\item[(x)] Finally let us remark that, if one of the glueballs
here considered contains an admixture of a $q\,\bar q$
state (indeed such admixtures are sometimes advocated
for in the theoretical literature, see e.g. \cite{clo}),
the cross section for $G\,\pi^0$ production might be
substantially increased since in this case the gluon--gluon
interaction would contribute here as well.
\end{description}

\acknowledgments

One of us (F.M.) wishes to thank the Laboratoire
de Physique Corpusculaire of the Coll\`ege de France,
especially its Director Prof. M.~Froissart and the
Theory Group, for the hospitality extended to him for six
weeks in summer 1993; he also acknowledges the support received
from the MURST (Ministero dell'Universit\`a e della Ricerca
Scientifica e Tecnologica, Italy).
%(Italian Ministry of University and of
%Scientific and Technological Research).

This work has been partially supported by the EU program
``Human Capital and Mobility'' under contract CHRX--CT94--0450.

\newpage

\appendix
\section{expressions of four--momenta, polarization
four--vectors and projectors of spinor pairs}

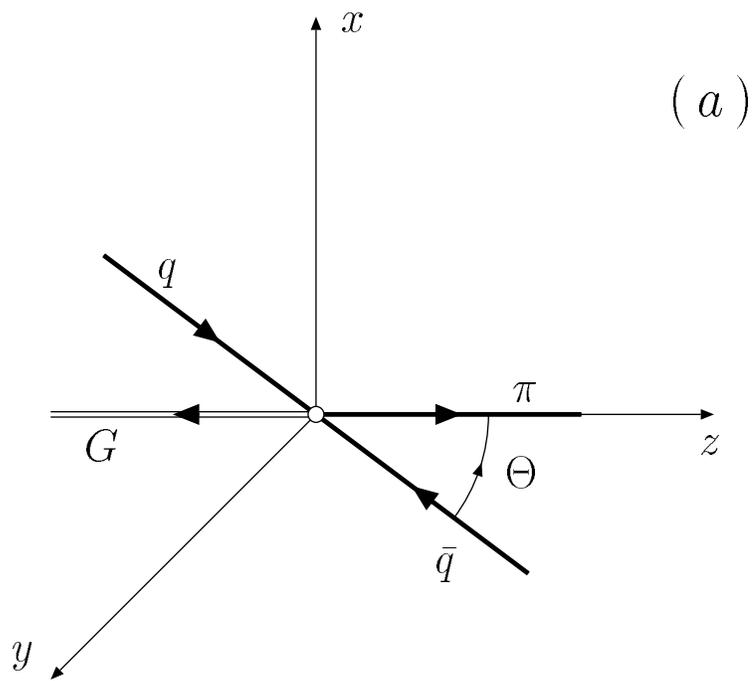
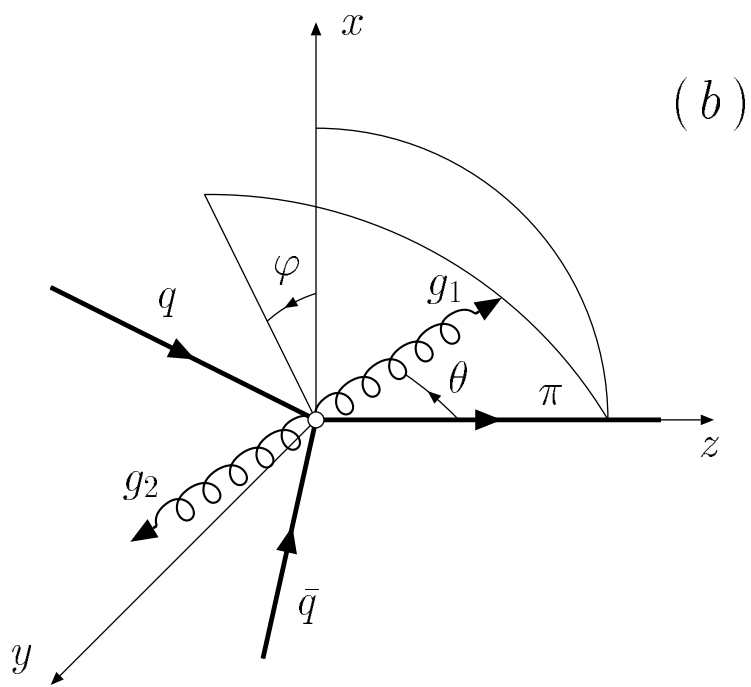
\begin{figure}
\caption[fig2]{ Kinematics schemes for: a) the process $q\,\bar q
\,\to\, G\,\pi$
in the center--of--mass frame of $q$ and $\bar q$;
b) the process $q\,\bar q \,\to\, g_1\,g_2\,\pi$ in the
center--of--mass frame of $g_1$ and $g_2$. }
\label{frame}
\end{figure}

For the four--momenta of initial partons and final particles
involved in the hard process, defining them in the
center--of--mass frame of that process (see Fig.~\ref{frame}a),
we use the following expressions
(components $0;x;y;z$ in that order):

\begin{eqnarray}
q^\mu & = & \frac{E}{2}\left( \begin{array}{c}
1 \\ -\sin\Theta \\ 0 \\ \cos\Theta \end{array} \right)
\qquad ; \qquad \bar q^\mu = \frac{E}{2}\left( \begin{array}{c}
1 \\ \sin\Theta \\ 0 \\ -\cos\Theta \end{array} \right) \\
G^\mu & = & \frac{E}{2}\left( \begin{array}{c}
1+\eta^2 \\ 0 \\ 0 \\ -1+\eta^2 \end{array} \right)
\qquad ; \qquad p_\pi^\mu = \frac{E}{2}(1-\eta^2)
\left( \begin{array}{c} 1 \\ 0 \\ 0 \\ 1 \end{array} \right)
\end{eqnarray}

%\begin{eqnarray}
%q & = & \frac{E}{2}(1,-\sin\Theta,0,\cos\Theta) \nonumber \\
%\bar q & = & \frac{E}{2}(1,\sin\Theta,0,-\cos\Theta) \nonumber \\
%G & = & \frac{E}{2}(1+\eta^2,0,0,-1+\eta^2) \\
%p_\pi & = & \frac{E}{2}(1-\eta^2)(1,0,0,1)
%\end{eqnarray}

\noindent while for the four--momenta of the intermediate
quarks $q'$, $\bar q'$ we simply have

\begin{equation}
q' = zp_\pi \qquad , \qquad \bar q' = (1-z)p_\pi
\end{equation}

Here we have called all four--momenta like the corresponding
particles (except for the pion), and we have defined:
$\eta = M/E$.

For the four--momenta of the intermediate gluons $g_1$,
$g_2$ and for their polarization four--vectors
$\epsilon^*_{1,\lambda_1}$, $\epsilon^*_{2,\lambda_2}$
($\lambda_{1,2} = 0,\pm 1$), we get, after performing
a Lorentz transformation from the glueball rest frame
(see Fig.~\ref{frame}b) to the c.m. frame of the
hard process ( Fig.~\ref{frame}a), the following
expressions:

\begin{equation}
g_1^\mu = \frac{E}{4}\left( \begin{array}{c}
1+\eta^2-(1-\eta^2)\beta\cos\theta \\
2\eta\beta\sin\theta\cos\varphi \\
2\eta\beta\sin\theta\sin\varphi \\
-1+\eta^2 + (1+\eta^2)\beta\cos\theta \end{array} \right)
\qquad ; \qquad g_2 = g_1(\beta \to -\beta)
\end{equation}

%\begin{eqnarray}
%g_1 & = & \frac{E}{4}(1+\eta^2-(1-\eta^2)\beta\cos\theta,
%2\eta\beta\sin\theta\cos\varphi,
%2\eta\beta\sin\theta\sin\varphi,
%(1+\eta^2)\beta\cos\theta-1+\eta^2) \nonumber \\
%g_2 & = & g_1(\beta \to -\beta)
%\end{eqnarray}

\begin{equation}
\epsilon^*_{1,\pm 1} = \epsilon^*_{2,\mp 1} = \mp \frac{1}{\sqrt{2}}
(\epsilon_{1x}\mp i\epsilon_{1y})
\end{equation}

\noindent with

\begin{equation}
\epsilon_{1x}^\mu =  \frac{1}{2\eta}\left( \begin{array}{c}
(1-\eta^2)\sin\theta \\ 2\eta\cos\theta\cos\varphi \\
2\eta\cos\theta\sin\varphi \\ -(1+\eta^2)\sin\theta
\end{array} \right)
\qquad ; \qquad \epsilon_{1y}^\mu = \left( \begin{array}{c}
0 \\ -\sin\varphi \\ \cos\varphi \\ 0 \end{array} \right)
\label{pol1}
\end{equation}

%\begin{eqnarray}
%\epsilon_{1x} & = & \frac{1}{2\eta}((1-\eta^2)\sin\theta,
%2\eta\cos\theta\cos\varphi,2\eta\cos\theta\sin\varphi,
%-(1+\eta^2)\sin\theta) \nonumber \\
%\epsilon_{1y} & = & (0, -\sin\varphi, \cos\varphi, 0)
%\end{eqnarray}

\noindent and finally

\begin{equation}
\epsilon_{1,0}^\mu = \frac{1}{\sqrt{1-\beta^2}}\frac{1}{2\eta}
\left( \begin{array}{c}
\beta(1+\eta^2)-(1-\eta^2)\cos\theta \\
2\eta\sin\theta\cos\varphi \\
2\eta\sin\theta\sin\varphi \\
-\beta(1-\eta^2)+(1+\eta^2)\cos\theta \end{array} \right)
\qquad ; \qquad
\epsilon_{2,0} = \epsilon_{1,0}(\beta \to -\beta)
\label{pol0}
\end{equation}

As for the projectors of the spinor pair corresponding
to $q'$, $\bar q'$, we make the substitution (in
accordance with Ref.~\cite{brod}, accounting for the
fact that the $(q',\bar q')$ system is in a pseudoscalar
state):

\begin{equation}
(v_{\bar q'}\bar u_{q'})_{PS} = \frac{1}{\sqrt{2}}
\bigl(v_{\bar q'}^\uparrow \bar u_{q'}^\downarrow -
v_{\bar q'}^\downarrow \bar u_{q'}^\uparrow \bigr) \,\to\,
\frac{1}{\sqrt{2}}\gamma_5 p_\pi\!\!\!\!\!\!\!\;/
\end{equation}

On the other hand, for the spinor pair corresponding to the
incoming quarks we use:

\begin{equation}
u_{q}^\uparrow \bar v_{\bar q}^\downarrow =
- \frac{E}{2\sqrt{2}}\,\epsilon_+\!\!\!\!\!\!/
\,\,\,\,(1-\gamma_5)
\end{equation}

\noindent with

\begin{equation}
\epsilon_+^\mu  = - \frac{1}{\sqrt{2}}
\left( \begin{array}{c}
0 \\ \cos\Theta \\ i \\ \sin\Theta \end{array} \right)
\end{equation}

\noindent noticing that no other helicity states of the incoming
quarks need to be considered in the calculation (see section
\ref{descri}).

As specified in section \ref{descri}, we let $\eta$ go to zero;
this is done, precisely, once we have computed the helicity
amplitudes of the subprocess $q\,\bar q \,\to\,
g_1\,g_2\,(q'\bar q')_{PS}$.
Then all divergences in $\eta$, due to the $\eta^{-1}$ factors
appearing in the expressions of the polarization four--vectors
of the gluons (see Eqs.~(\ref{pol1}),(\ref{pol0})) must vanish.
This (as well as the vanishing of all divergences in $\beta$
after applying Eq.~(\ref{bound})) is a good check of the
correctness of the calculation.

\section{helicity amplitudes for the processes \protect\\
\protect$\bbox{\lowercase{q\,\bar q} \,\to\,
G\,(\lowercase{q'\bar q'})_{PS}}$ and
\protect$\bbox{\lowercase{q\,\bar q} \,\to\, G\,\pi}$}

In this appendix we present the expressions of
all independent helicity amplitudes for the
process $q\,\bar q \,\to\, G\,(q'\bar q')_{PS}$, {\it i.e.}
${\cal M}^{\lambda\,\bar\lambda,\,\Lambda}_{q\,\bar q \,\to\,
G\,(q'\bar q')_{PS}}(E,\Theta,z)$, before their convolution with the
pion distribution amplitude (see Eq.~(\ref{conv})), but
with the final quark--antiquark pair being specified to be
in a pseudoscalar spin state.
In each case, the corresponding helicity amplitudes
for the process $q\,\bar q \,\to\, G\,\pi$, {\it i.e.}
${\cal M}^{\lambda\,\bar\lambda,\,\Lambda}_{q\,\bar q \,\to\,
G\,\pi}(E,\Theta)$, after convolution with the
asymptotic resp. Chernyak--Zhitnitsky pion distribution
amplitude are also given. All helicity amplitudes
not explicitly shown here can be derived by means of
symmetry properties (see section II) from those given hereafter,
or are vanishing.
We have fixed the helicities of the initial quark--antiquark
pair as follows: $\lambda=1/2$, $\bar\lambda=-1/2$.
For shortness, we use the following notation:
${\cal M}^\Lambda_{PS}$ for ${\cal M}^{\lambda\,\bar\lambda,\,\Lambda}
_{q\,\bar q \,\to\, G\,(q'\bar q')_{PS}}(E,\Theta,z)$, and
${\cal M}^\Lambda_{(as)}$ resp. ${\cal M}^\Lambda_{(CZ)}$
for the amplitudes obtained after convolution of
${\cal M}^\Lambda_{PS}$ with the asymptotic resp.
the Chernyak--Zhitnitsky pion distribution amplitude.
Furthermore we use $c=\cos\Theta$, $s=\sin\Theta$,
$u=z(1-z)$, $w=2z-1$,
${\cal L}_c = \ln[(1+\cos\Theta)/(1-\cos\Theta)]$.
The constant $f_L$ has been defined in Eq.~(\ref{fl}).
Finally, as usual, $g_s = \sqrt{4\pi\alpha_s}$, while
$i$, $j$ are the color indices of the incoming
quark and antiquark respectively.

\begin{description}

\item[(i)] $L=S=J=0$
\begin{mathletters}
\begin{eqnarray}
{\cal M}^0_{PS} & = & -\frac{8}{27}g_s^4 f_0 \delta_{ij}
\frac{1}{E^2 s^3}\frac{-8s^4+s^2(-96u+25)+64u}{u(1-cw)}  \\
{\cal M}^0_{(as)} & = & -\frac{8}{9\sqrt{3}} g_s^4 f_0 f_\pi
\delta_{ij} \frac{1}{E^2s^3}
\Bigl\{ 24 - \frac{8}{c^2} + \frac{1-c^2}{2c^3}
\bigl(8-7c^2+8c^4\bigr){\cal L}_c\Bigr\} \\
{\cal M}^0_{(CZ)} & = & -\frac{40}{9\sqrt{3}} g_s^4 f_0 f_\pi
\delta_{ij} \frac{1}{E^2s^3}
\Bigl\{8c^2-7+\frac{37}{3c^2}-\frac{8}{c^4}+\frac{1-c^2}{2c^5}
\bigl(8-7c^2+8c^4\bigr){\cal L}_c\Bigr\}
\end{eqnarray}
\end{mathletters}
\item[(ii)] $L=S=1$, $J=0$
\begin{mathletters}
\begin{eqnarray}
{\cal M}^0_{PS} & = & -\frac{8\sqrt{2}}{81}g_s^4 f_1
\delta_{ij} \frac{1}{E^2s^3}\frac{1}{u(1-cw)^2} \nonumber \\
& \times & \Bigl\{s^4(4c+5w)+4s^2\bigl[c(21u-4)+
4w(5u+1)\bigr]+64u(c-w)\Bigr\} \\
{\cal M}^0_{(as)} & = & -\frac{8\sqrt{2}}{27\sqrt{3}}
g_s^4 f_1 f_\pi \delta_{ij} \frac{1}{E^2s^3}
\Bigl\{-4c^3+25c+\frac{7}{c}-\frac{12}{c^3} \nonumber \\
& & \quad\quad\quad\quad\quad
+\ \frac{1-c^2}{2c^4}\bigl(12+c^2+5c^4\bigr){\cal L}_c \Bigr\} \\
{\cal M}^0_{(CZ)} & = & -\frac{40\sqrt{2}}{27\sqrt{3}}
g_s^4 f_1 f_\pi \delta_{ij} \frac{1}{E^2s^3}
\Bigl\{4c^3+\frac{1}{c}+\frac{61}{3c^3}-\frac{20}{c^5}
\nonumber \\ & & \quad\quad\quad\quad\quad +\
\frac{1-c^2}{2c^6}\bigl(20-7c^2-3c^4+8c^6\bigr)
{\cal L}_c \Bigr\}
\end{eqnarray}
\end{mathletters}
\item[(iii)] $L=0$, $S=2$, $J=2$
\begin{mathletters}
\begin{eqnarray}
{\cal M}^{\pm 2}_{PS} & = & \frac{512}{9\sqrt{3}} g_s^4 f_0
\delta_{ij} \frac{1}{E^2 s(1\pm c)}\frac{1}{(1\mp w)} \\
{\cal M}^{\pm 2}_{(as)} & = & \frac{128}{9} g_s^4 f_0 f_\pi
\delta_{ij} \frac{1}{E^2s(1\pm c)} \\
{\cal M}^{\pm 2}_{(CZ)} & = & \frac{5}{3}{\cal M}^{\pm 2}_{(as)}
\quad\quad\quad\quad\quad\quad\quad\quad\quad\quad\quad\quad
\quad\quad\quad\quad\quad\quad\quad\quad\quad\quad\quad\quad
\end{eqnarray}
\end{mathletters}
\begin{mathletters}
\begin{eqnarray}
{\cal M}^{0}_{PS} & = & -\frac{8\sqrt{2}}{27}g_s^4 f_0
\delta_{ij} \frac{1}{E^2s^3}\frac{-4s^4-s^2(48u+1)+32u}{u(1-cw)} \\
{\cal M}^{0}_{(as)} & = & -\frac{8\sqrt{2}}{9\sqrt{3}}
g_s^4 f_0 f_\pi \delta_{ij} \frac{1}{E^2s^3}
\Bigl\{12-\frac{4}{c^2}+\frac{1-c^2}{2c^3}(4-c^2)
(1-4c^2){\cal L}_c \Bigr\} \\
{\cal M}^{0}_{(CZ)} & = & -\frac{40\sqrt{2}}{9\sqrt{3}}
g_s^4 f_0 f_\pi \delta_{ij} \frac{1}{E^2s^3}
\Bigl\{4c^2-17+\frac{59}{3c^2}-\frac{4}{c^4}
\nonumber \\ & & \quad\quad\quad\quad\quad +\
\frac{1-c^2}{2c^5}(4-c^2)(1-4c^2){\cal L}_c \Bigr\}
\end{eqnarray}
\end{mathletters}
\item[(iv)] $L=2$, $S=0$, $J=2$
\begin{mathletters}
\begin{eqnarray}
{\cal M}^{\pm 2}_{PS} & = & \frac{1024\sqrt{2}}{135\sqrt{3}}
g_s^4 f_2 \delta_{ij} \frac{1}{E^2 s(1\pm c)}\frac{1}{1\mp w} \\
{\cal M}^{\pm 2}_{(as)} & = & \frac{256\sqrt{2}}{135}
g_s^4 f_2 f_\pi \delta_{ij} \frac{1}{E^2s(1\pm c)} \\
{\cal M}^{\pm 2}_{(CZ)} & = & \frac{5}{3}{\cal M}^{\pm 2}_{(as)}
\quad\quad\quad\quad\quad\quad\quad\quad\quad\quad\quad\quad
\quad\quad\quad\quad\quad\quad\quad\quad\quad\quad\quad\quad
\end{eqnarray}
\end{mathletters}
\begin{mathletters}
\begin{eqnarray}
{\cal M}^{0}_{PS} & = & -\frac{32}{405} g_s^4 f_2
\delta_{ij} \frac{1}{E^2s^3}\frac{1}{u(1-cw)^3} \nonumber \\
&\times& \Bigl\{-8s^6(2u-1)+s^4\bigl[7cw+2u(-224u+137)-40\bigr]
\nonumber \\
& & +s^2\bigl[22cw(10u-3)+448u(2u-1)+66\bigr]-192u(cw+2u-1)\Bigr\} \\
{\cal M}^{0}_{(as)} & = & -\frac{32}{135\sqrt{3}}
g_s^4 f_2 f_\pi \delta_{ij} \frac{1}{E^2s^3}
\Bigl\{-5c^2+\frac{63}{2}+\frac{43}{2c^2} \nonumber \\
& & \quad\quad\quad\quad\quad +
\ \frac{1-c^2}{4c^5}\bigl(48-11c^2+9c^4+8c^6\bigr)
{\cal L}_c \Bigr\} \\
{\cal M}^{0}_{(CZ)} & = & -\frac{32}{27\sqrt{3}}
g_s^4 f_2 f_\pi \delta_{ij} \frac{1}{E^2s^3}
\Bigl\{-3c^2+\frac{193}{6}-\frac{373}{6c^2}
+\frac{101}{c^4} -\frac{60}{c^6} \nonumber \\
& & \quad\quad\quad\quad\quad +
\ \frac{1-c^2}{4c^7}\bigl(120-122c^2+59c^4-
11c^6+8c^8\bigr){\cal L}_c \Bigr\}
\end{eqnarray}
\end{mathletters}
\item[(v)] $L = S = J = 2$
\begin{mathletters}
\begin{eqnarray}
{\cal M}^{\pm 2}_{PS} & = & \frac{256\sqrt{2}}{135\sqrt{21}}
g_s^4 f_2 \delta_{ij} \frac{1}{E^2 s(1\pm c)}\frac{1}{1\mp w} \\
{\cal M}^{\pm 2}_{(as)} & = & \frac{64\sqrt{2}}{135\sqrt{7}}
g_s^4 f_2 f_\pi \delta_{ij} \frac{1}{E^2 s(1\pm c)} \\
{\cal M}^{\pm 2}_{(CZ)} & = & \frac{5}{3}{\cal M}^{\pm 2}_{(as)}
\quad\quad\quad\quad\quad\quad\quad\quad\quad\quad\quad\quad
\quad\quad\quad\quad\quad\quad\quad\quad\quad\quad\quad\quad
\end{eqnarray}
\end{mathletters}
\begin{mathletters}
\begin{eqnarray}
{\cal M}^{0}_{PS} & = & \frac{8}{405\sqrt{7}}g_s^4 f_2
\delta_{ij} \frac{1}{E^2s^3}\frac{1}{u(1-cw)^3}
\Bigl\{4s^6(68u-13)\nonumber \\ & + &
s^4\bigl[-140cw+4u(560u-311)+311\bigr]
+ 2s^2\bigl[-cw(568u-171) \nonumber \\
& + & 2u(-784u+527)-171\bigr]+576u(cw+2u-1)\Bigr\} \\
{\cal M}^{0}_{(as)} & = & \frac{8\sqrt{3}}{405\sqrt{7}}
g_s^4 f_2 f_\pi \delta_{ij} \frac{1}{E^2s^3}
\Bigl\{-20c^2-180+\frac{224}{c^2}-\frac{96}{c^4} \nonumber \\
& & \quad\quad\quad\quad +\ \frac{1-c^2}{2c^5}
\bigl(96-160c^2+105c^4-68c^6\bigr){\cal L}_c \Bigr\} \\
{\cal M}^{0}_{(CZ)} & = & \frac{8\sqrt{3}}{81\sqrt{7}}
g_s^4 f_2 f_\pi \delta_{ij} \frac{1}{E^2s^3}
\Bigl\{-96c^2+\frac{1295}{3}-\frac{2795}{3c^2}
+\frac{812}{c^4}-\frac{240}{c^6} \nonumber \\
& & \quad\quad\quad\quad +\ \frac{1-c^2}{2c^7}
\bigl(240-652c^2+529c^4-160c^6+16c^8\bigr){\cal L}_c \Bigr\}
\end{eqnarray}
\end{mathletters}
\item[(vi)] $L=S=2$, $J=4$
\begin{mathletters}
\begin{eqnarray}
{\cal M}^{\pm 2}_{PS} & = & \frac{1024\sqrt{2}}{135\sqrt{7}}
g_s^4 f_2 \delta_{ij} \frac{1}{E^2 s(1\pm c)}\frac{1}{1 \mp w} \\
{\cal M}^{\pm 2}_{(as)} & = & \frac{256\sqrt{2}}{45\sqrt{21}}
g_s^4 f_2 f_\pi \delta_{ij} \frac{1}{E^2 s(1\pm c)} \\
{\cal M}^{\pm 2}_{(CZ)} & = & \frac{5}{3}{\cal M}^{\pm 2}_{(as)}
\quad\quad\quad\quad\quad\quad\quad\quad\quad\quad\quad\quad
\quad\quad\quad\quad\quad\quad\quad\quad\quad\quad\quad\quad
\quad\quad
\end{eqnarray}
\end{mathletters}
\begin{mathletters}
\begin{eqnarray}
{\cal M}^{0}_{PS} & = & \frac{32}{135\sqrt{35}}
g_s^4 f_2 \delta_{ij} \frac{1}{E^2s^3}\frac{1}{u(1-cw)^3}
\nonumber \\
&\times& \Bigl\{8s^6(2u-1)+s^4\bigl[-7cw+64u(7u-4)+22\bigr]
\nonumber \\
& & +s^2\bigl[10cw(-22u+3)+8u(-112u+47)-30\bigr]
+192u(cw+2u-1)\Bigr\} \\
{\cal M}^{0}_{(as)} & = & \frac{32\sqrt{3}}{135\sqrt{35}}
g_s^4 f_2 f_\pi \delta_{ij} \frac{1}{E^2s^3}
\Bigl\{5c^2-18-\frac{35}{c^2}+\frac{24}{c^4} \nonumber \\
& & \quad\quad\quad\quad\quad -\ \frac{1-c^2}{2c^5}\bigl(24-19c^2+4c^6)
{\cal L}_c \Bigr\} \\
{\cal M}^{0}_{(CZ)} & = & \frac{32\sqrt{3}}{27\sqrt{35}}
g_s^4 f_2 f_\pi \delta_{ij} f_\pi \frac{1}{E^2s^3}
\Bigl\{3c^2-\frac{245}{3}+\frac{578}{3c^2}
-\frac{182}{c^4}+\frac{60}{c^6} \nonumber \\
& & \quad\quad\quad\quad\quad -\
\frac{1-c^2}{2c^7}\bigl(60-142c^2+106c^4-19c^6+4c^8\bigr)
{\cal L}_c \Bigr\}
\end{eqnarray}
\end{mathletters}

\end{description}

\end{document}